\documentclass[aps,pra,superscriptaddress,showpacs,floatfix,twocolumn,citeautoscript]{revtex4-1}

\usepackage{amssymb}
\usepackage{amsmath}
\usepackage{mathrsfs}
\usepackage{graphicx}
\usepackage{bm}
\usepackage{braket}
\usepackage{float}
\usepackage{color}
\usepackage{times}
\usepackage{MnSymbol}
\usepackage{verbatim}
\usepackage{enumerate}

\newcommand{\bea}{\begin{eqnarray}}
\newcommand{\eea}{\end{eqnarray}}
\newcommand{\be}{\begin{equation}}
\newcommand{\ee}{\end{equation}}

\newcommand{\ci}{\mathrm{i}}

\newcommand{\xv}{\hat{\bm{x}}}
\newcommand{\yv}{\hat{\bm{y}}}

\begin{document} 

\title{Multi-orbital tight binding model for cavity-polariton lattices}

\author{Franco Mangussi}
\affiliation{Centro At{\'{o}}mico Bariloche and Instituto Balseiro,
Comisi\'on Nacional de Energ\'{\i}a At\'omica (CNEA)--Universidad Nacional de Cuyo (UNCUYO), 8400 Bariloche, Argentina}
\affiliation{Instituto de Nanociencia y Nanotecnolog\'{i}a (INN), Consejo Nacional de Investigaciones Cient\'{\i}ficas y T\'ecnicas (CONICET)--CNEA, 8400 Bariloche, Argentina}
\author{Marijana Mili\'cevi\'c}
\affiliation{Centre de Nanosciences et de Nanotechnologies (C2N), CNRS, Universit\'{e} Paris-Sud, Universit\'{e} Paris-Saclay, 91120 Palaiseau, France}
\author{Isabelle~Sagnes}
\affiliation{Centre de Nanosciences et de Nanotechnologies (C2N), CNRS, Universit\'{e} Paris-Sud, Universit\'{e} Paris-Saclay, 91120 Palaiseau, France}
\author{Luc~Le~Gratiet}
\affiliation{Centre de Nanosciences et de Nanotechnologies (C2N), CNRS, Universit\'{e} Paris-Sud, Universit\'{e} Paris-Saclay, 91120 Palaiseau, France}
\author{Abdelmounaim~Harouri}
\affiliation{Centre de Nanosciences et de Nanotechnologies (C2N), CNRS, Universit\'{e} Paris-Sud, Universit\'{e} Paris-Saclay, 91120 Palaiseau, France}
\author{Aristide~Lema\^itre}
\affiliation{Centre de Nanosciences et de Nanotechnologies (C2N), CNRS, Universit\'{e} Paris-Sud, Universit\'{e} Paris-Saclay, 91120 Palaiseau, France}
\author{Jacqueline~Bloch}
\affiliation{Centre de Nanosciences et de Nanotechnologies (C2N), CNRS, Universit\'{e} Paris-Sud, Universit\'{e} Paris-Saclay, 91120 Palaiseau, France}
\author{Alberto~Amo}
\affiliation{Universit\'e de Lille, CNRS, UMR 8523 --PhLAM-- Physique des Lasers Atomes et Mol\'ecules, F-59000 Lille, France}
\author{Gonzalo~Usaj}
\affiliation{Centro At{\'{o}}mico Bariloche and Instituto Balseiro,
Comisi\'on Nacional de Energ\'{\i}a At\'omica (CNEA)--Universidad Nacional de Cuyo (UNCUYO), 8400 Bariloche, Argentina}
\affiliation{Instituto de Nanociencia y Nanotecnolog\'{i}a (INN), Consejo Nacional de Investigaciones Cient\'{\i}ficas y T\'ecnicas (CONICET)--CNEA, 8400 Bariloche, Argentina}

\begin{abstract}
In this work we present a tight-binding model that allows to describe with a minimal amount of parameters the band structure of exciton-polariton lattices. This model based on $s$ and $p$ non-orthogonal photonic orbitals faithfully reproduces experimental results reported for polariton graphene ribbons. We analyze in particular the influence of the non-orthogonality, the inter-orbitals interaction and the photonic spin-orbit coupling on the polarization and dispersion of bulk bands and edge states. 
\end{abstract}
\date{\today}
\maketitle
\section{Introduction \label{Introduction}}

Coupled photonic resonators have appeared in the past few years as an excellent platform to engineer lattice Hamiltonians~\cite{Nolte2010, Houck2012, Bellec2013, Jacqmin2014}. The possibility of controlling the geometry, on-site energy and hopping along with the spectroscopic access to the momentum- and real-space distributions of the wavefunctions 
are opening new perspectives in the study of elaborate solid-state Hamiltonians in the photonics realm. In addition, the engineering of gain and losses and the presence of Kerr nonlinearities are unveiling genuinely photonic phenomena in lattices, which include lasing in topological states~\cite{St-Jean2017,Bahari2017,Bandres2018,Zhao2018,Parto2018,Klembt2018}, PT-symmetric~\cite{Weimann2016} and charge conjugated phases~\cite{Poli2015} and the observation of dissipative phase transitions~\cite{Fitzpatrick2017,Rodriguez2017}.

Lattices of polariton resonators in semiconductor microcavities provide one of the most versatile platforms to implement this kind of Hamiltonians~\cite{Schneider2017}. Polaritons are hybrid light-matter quasiparticles that arise from the strong coupling of quantum-well excitons and photons confined in a micron-scale Fabry-Perot cavity. Their excitonic component results in significant polariton interactions and in sensitivity to external magnetic fields. The first feature has allowed the observation of bi-stability~\cite{Baas2004}, polariton superfluidity~\cite{Amo2009} and solitons~\cite{Amo2011,Sich2011} in planar structures, while the second has been used to demonstrate lasing in circularly polarised states~\cite{Sturm2015} and in chiral edge states~\cite{Klembt2018}.

A very convenient way to implement lattices of polariton resonators is by confining their photonic component in fully or partially etched structures. The building block of these lattices is typically a resonator of cylindrical symmetry, in which photons are confined in the three spatial directions. The polariton resonators show confined modes separated by a gap, each of them with a particular geometry: the ground state is formed by cylindrically symmetric $s$-modes, the first excited state is doubly degenerate with $p$-type modes, the next states have $d$-symmetry and so on. Such confined modes have been realized by fully etching the semiconductor structure~\cite{Bayer1998,Bajoni2008}, by partial etching of the upper cavity mirror~\cite{Klembt2017,Whittaker2018}, by growth interruption and etching of the cavity spacer~\cite{Kaitouni2006,Winkler2015}, and in half cavities closed by an external mirror~\cite{Zhang2014,Dufferwiel2014,Besga2015}. By laterally coupling the photonic modes of the resonators, lattices of different geometries have been implemented, including one-dimensional regular~\cite{Bayer1999,Tanese2013,Winkler2016}, Stub~\cite{Baboux2016}, Su-Schrieffer-Heeger (SSH) lattices~\cite{St-Jean2017} and aperiodic lattices~\cite{Tanese2014}, and two-dimensional honeycomb~\cite{Jacqmin2014,Klembt2018} and Lieb lattices~\cite{Klembt2017,Whittaker2018}, showing a wide variety of dispersions and topological features. One of the great assets of this system is the possibility of designing lattices with synthetic strain, which have been recently employed to engineer new types of Dirac cones~\cite{Milicevic2018} and are promising to engineer artificial gauge fields~\cite{Rechtsman2013,Salerno2015}.

The design of polariton lattices and the interpretation of the polariton bands measured in photoluminescence studies have so far largely relied on the mapping to a tight-binding model. In this model, each orbital mode of each cylindrical microresonator is independent from the other orbitals and plays the role of a point-like tight-binding site, all of them with identical
on-site energy, coupled to their nearest neighbors. In SSH, Lieb and honeycomb geometries, this kind of tight-binding Hamiltonian presents chiral symmetry and, therefore, the upper and lower bands of eigenvalues should be mirror symmetric with respect to the value of the on-site energy. However, this simple model shows significant deviations from the experimentally observed dispersions, both in 1D and in 2D lattices~\cite{Jacqmin2014,Klembt2018,Winkler2016,Baboux2016,Klembt2017,Whittaker2018}. In particular, in experimental observations, a significant asymmetry between upper and lower bands is systematically observed. An efficient way to fit this band asymmetry is to add a next-nearest neighbor coupling to the tight-binding model. This technique was used, for instance, in the works of Jacqmin, Baboux and coworkers~\cite{Jacqmin2014,Baboux2016}.

Despite the apparent success of the fits, the question of the physical relevance of the actual next-nearest neighbor coupling remains, particularly in structures based on complete etching of the semiconductor microcavities, for which the photonic confinement is expected to be very strong within the physical dimensions of the micropillar. Therefore, the observed band asymmetries call for other corrections to the tight binding description. One of them is the coupling between modes of different symmetry belonging to nearest neighbor sites: $s$ and $p$-modes or $p$ and $d$-modes in adjacent micropillars. Simultaneously, the significant spatial overlap between adjacent micropillars in real structures raises questions about the accuracy of the tight binding model, which assumes the limit of weak overlaps. When the overlaps are significant, the original basis made of the individual uncoupled resonators is far from an orthogonal basis, and non-orthogonal corrections need to be added to the original tight-binding Hamiltonian. In models like the honeycomb lattice, these corrections have been shown to result in band asymmetries quite similar to those induced by next-nearest neighbors~\cite{McKinnon1995}. Understanding the effects of these corrections is of crucial importance to interpret a number of physical observations within this model. 

In this article, we show that, indeed, the experimental dispersion of lattices of polariton micro-pillars can be described with very high accuracy using a realistic tight-binding model that takes into account both the non-orthogonality of the micro-pillar basis and the coupling between $s$- and $p$-bands. We show in this way that direct next-nearest neighbors coupling is not necessary to fully reproduce all experimentally observed phenomenology. To complete our description we take into account the TE-TM splitting characteristic of dielectric microcavities. We compare our model to experimental dispersions obtained in a honeycomb lattice of coupled micro-pillars. Our results should improve significantly polariton tight-binding models.

The rest of the paper is organized as follows: in Section \ref{Model} we introduce the basics of our non-orthogonal tight binding model and a simple variational approach based on low contrast refraction indices that implements an effective modeling of the single pillar photonic modes inside the lattice. We apply our model to the case of a honeycomb lattice in Section \ref{Results} and compare our results with experimental data showing that they agree quite well, even in the case of distorted lattices. Finally, we conclude in Section \ref{conclu}.   
\section{A minimal tight-binding description for cavity-polariton lattices\label{Model}}
In this Section we present a simple tight-binding (TB) approach to describe cavity-polariton lattices made out of single cavity micropillars with several polaritonic modes of different symmetries. At the core of the method lies the fact that we will consider the case of weakly coupled cavities where the photonic modes of a single cavity are a good starting point of the calculation. We will explicitly take into account the overlap between photonic modes at nearest neighbors cavities, including those with different symmetries, since this turns out to be very important to describe the experimental data. Our approach is similar in spirit to the one developed in Ref.~\cite{Kamalakis2006} for photonic crystals, where the global photonic field was written as a linear combination of the modes corresponding to isolated pillars located at each lattice site. Here, however, we will show that to effectively capture the behavior of the real photonic modes of a pillar due to the spatial overlap with its neighbors it is important not to consider the modes of an isolated pillar in vacuum but those of a pillar surrounded by an effective media. 
\subsection{Non-orthogonal tight-binding approach \label{Non-orthogonal tight-binding approach}}
We first summarize the basics of the usual tight-binding (TB) approach involving a non-orthogonal set of localized orbitals in a lattice (see for instance Refs.~\cite{McKinnon1995} and \cite{Paxton2009}). For simpli\-city, we start by considering the case of a single orbital per site. Generalization to multi-orbital sites is done at the end the section. 

For a system with $N$ sites positioned at $\bm{R}_{i}$ with $i=1,...,N$, the single-particle wave-function in the TB approximation is given by the linear combination, 
\begin{equation}
\label{1}
 \ket{\Psi}=\sum_{i=1}^{N} c_{i} \ket{\psi_{i}}\,.
\end{equation}
Here $\ket{\psi_{i}}$ is the orbital state of a single pillar at site $\bm{R}_{i}$ which is assumed to be normalized. The Schr\"{o}dinger equation $\hat{H}\ket{\Psi}=\varepsilon \ket{\Psi}$ can then be reduced to the following matrix equation
\begin{equation}
\label{2}
 \bm{H}\bm{c}=\varepsilon\, \bm{S}\bm{ c}\,,
\end{equation}
where we have introduced the notation $\bm{c}=(c_1,c_2,\dots,c_N)^\mathrm{T}$. The elements of the Hamiltonian ($\bm{H}$) and overlap ($\bm{S}$) matrices are given by 
\begin{equation}
\label{3}
H_{i j}=\bra{\psi_{i}}\hat{H} \ket{\psi_{j}}\,,\qquad
S_{i j}=\braket{\psi_{i}|\psi_{j}}\,.
\end{equation}
In the case of an orthogonal basis, $H_{i j}$ is simply either the on-site energy $\varepsilon_i$ of the orbital $\ket{\psi_i}$ (for $i=j$) or the so-called hopping matrix element $t_{i j}$ (for $i\neq j$) that comes from the inter-site potential term. Both are usually taken as independent parameters. In the non-orthogonal case however, these terms are mixed and $H_{i j}$ can be parameterized in different (equivalent) ways.  We choose the following one
\begin{equation}
\label{3b}
H_{i j}=\frac{\left(\varepsilon_{i}+\varepsilon_{j}\right)}{2} S_{i j}+t_{i j}\,,
\end{equation}
with $t_{ii}=0$. This symmetric way to represent $H_{i j}$ will allow us later on, to make a simple approximation to the  inter-orbital coupling ($t_{ij}$) and to consider the cases where the on-site energy changes from site to site. Notice also that this parametrization explicitly takes into account that a global shift of the  site energies translates in a global shift of the bands. On the other hand, since in our case the  orbital states are taken to be real functions we have that $S_{ij}=S_{ji}$.
Equation (\ref{2}) can be solved by making the substitution $\bm{c}=\bm{S}^{-\frac{1}{2}} \tilde{\bm{c}}$ to get the more familiar orthogonal eigenvalue problem
\begin{equation}
\label{4}
\tilde{\bm{H}} \tilde{\bm{c}}=\varepsilon\, \tilde{\bm{c}}\, ,
\end{equation}   
with 
\begin{equation}
\label{5}
\tilde{\bm{H}}=\bm{S}^{-\frac{1}{2}} \bm{H} \bm{S}^{-\frac{1}{2}}\, .
\end{equation}   
Once solved, $\bm{c}$ can be recovered from $\tilde{\bm{c}}$ by back-substitution.
A very convenient way of writing the Hamiltonian is in terms of se\-cond quantization operators. Hence, we introduce a set of bosonic creation and annihilation operators, $\hat{b}_{i}^{\dagger}$ and $\hat{b}_{i}$, respectively, which allow us to rewrite the Hamiltonian as
\begin{equation}
\label{7}
\hat{H}=\sum_{i j} \left[\frac{\left(\varepsilon_{i}+\varepsilon_{j}\right)}{2} S_{i j}+t_{i j}\right] \hat{b}_{i}^{\dagger} \hat{b}_{j}^{}.
\end{equation}   
These operators, in the non-orthogonal case, are not those that create or annihilate a particle in the state $ \ket{\psi_{i}}$. In fact, they are a linear combination of the latter. If we  denote such operators by $\hat{a}_{i}^{\dagger}$, so that $\ket{\psi_{i}}=\hat{a}_{i}^{\dagger} \ket{0}$ where $\ket{0}$ is the vacuum state, then  we have that 
\begin{equation}
\label{8}
\hat{b}_{i}^{\dagger}=\sum_{j} S_ {i j}^{-1}\hat{a}_{j}^{\dagger}\,,
\end{equation}  
with 
$\left[\hat{a}_{i}^{\dagger},\hat{a}_{j}\right] = S_{ij}$
and
$\left[\hat{a}_{i}^{\dagger},\hat{a}_{j}^{\dagger}\right]= \big[\hat{a}_{i}^{},\hat{a}_{j}\big]=0$.

\subsection{Multi-orbital model \label{Multi-orbital model}}
\begin{figure}[t]
\centering
\includegraphics[width=0.95\columnwidth]{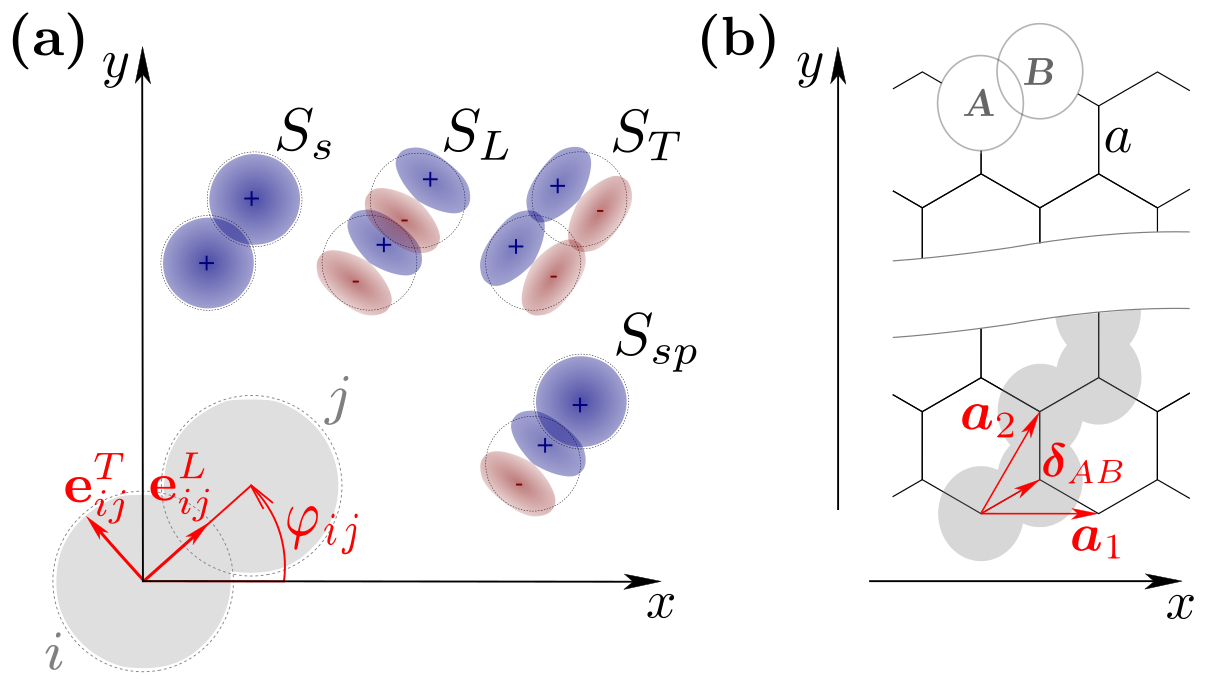}
\caption{(Color online) (a) Sketch of two adjacent micropillars, the corresponding low energy photonic modes and their overlaps: $s$ orbitals ($S_{s}$), $p$ orbitals along the bond ($S_{L}$) and transversal to it ($S_{T}$) and the coupling between $s$ and $p$ orbitals ($S_{sp}$). (b) Scheme of a polaritonic honeycomb lattice and a choice for the corresponding lattice vectors. Here $a$ is the distance between nearest neighbors.}
\label{schemes1}
\end{figure} 
Let us now consider the more realistic case of a lattice in which each micropillar supports several polariton modes of different symmetry related to its polar angle distribution. We restrict ourselves to the case of $s$ and $p$ orbitals because it will be enough to explain the experimental data described below. Generalization to more orbitals is straightforward by applying the same procedure.
Using the non-orthogonal tight-binding approximation described above, and considering only nearest-neighbors (NNs) overlap and hopping terms, the Hamiltonian (\ref{7}) can be written  as 
\begin{equation}
\label{H}
\hat{H}=\hat{H}_{s}+\hat{H}_{p}+\hat{H}_{sp}\, ,
\end{equation}
where the first two terms describe the coupling between the same type of orbitals ($s$ and $p$), and the last one the coupling between different types of orbitals ($s-p$ coupling). In its most general form, and accounting for the two polarization modes of each orbital, these three terms are given by
\begin{widetext}
\begin{eqnarray}
\label{10}
\hat{H}_{s}&=&\sum_{i,\sigma} \varepsilon_{i s \sigma} \, \hat{b}_{i s \sigma}^{\dagger}\hat{b}_{i s \sigma}^{\,}+ \sum_{\braket{i,j},\sigma} \left(\frac{ \left(\varepsilon_{i s \sigma}+\varepsilon_{j s \sigma}\right)}{2}S_{s}+t_s \right) \left(\hat{b}_{i s \sigma}^{\dagger} \hat{b}_{j s \sigma}^{\,}                                           +\Delta \, e^{-2 \ci \varphi_{ij} \sigma} \, \hat{b}_{i s \sigma}^{\dagger} \hat{b}_{j s \bar{\sigma}}^{\,}\right)\,,
\\ 
\nonumber
\hat{H}_{p}&=&\sum_{i,\sigma} \varepsilon_{i p \sigma} \, \hat{\bm{b}}_{i p \sigma}^{\, \dagger} \cdot \, \hat{\bm{b}}_{i p \sigma}
\\ \nonumber
&+&  \sum_{\braket{i,j},\sigma} \left(\frac{\left(\varepsilon_{i p \sigma}+\varepsilon_{j p \sigma}\right)}{2}S_{L}+t_L\right)\left[\left(\hat{\bm{b}}_{i p \sigma}^{\, \dagger} \cdot \, \bm{e}_{i j}^{\,L} \right) \left(\bm{e}_{i j}^{\, L} \cdot \hat{\bm{b}}_{j p \sigma}\right)+\Delta \, e^{-2 \ci \varphi_{ij}\sigma}\left(\hat{\bm{b}}_{i p \sigma}^{\, \dagger} \cdot \, \bm{e}_{i j}^{\,L} \right) \left(\bm{e}_{i j}^{\,L} \cdot \hat{\bm{b}}_{j p \bar{\sigma}}\right)\right]
\\ 
\label{11}
&+& \sum_{\braket{i,j},\sigma}\left(\frac{\left(\varepsilon_{i p \sigma}+\varepsilon_{j p \sigma}\right)}{2}S_{T} +t_T\right)\left[\left(\hat{\bm{b}}_{i p \sigma}^{\, \dagger} \cdot \, \bm{e}_{i j}^{\,T} \right) \left(\bm{e}_{i j}^{\,T} \cdot \hat{\bm{b}}_{j p \sigma}\right)+\Delta \, e^{-2 \ci \varphi_{ij}\sigma} \left(\hat{\bm{b}}_{i p \sigma}^{\, \dagger} \cdot \, \bm{e}_{i j}^{\,T} \right) \left(\bm{e}_{i j}^{\, T} \cdot \hat{\bm{b}}_{j p \bar{\sigma}}\right)\right]\,,
\\
\label{12}                                                 
\hat{H}_{sp}&=& \sum_{\braket{i,j},\sigma}\left(\frac{\left(\varepsilon_{i s \sigma}+\varepsilon_{j p \sigma}\right)}{2}S_{sp} +t_{sp} \right)\left[\hat{b}_{i s \sigma}^{\dagger} \left(\bm{e}_{i j}^{\,L} \cdot \hat{\bm{b}}_{j p \sigma}\right)+\Delta \, e^{-2 \ci \varphi_{ij} \sigma} \, \hat{b}_{i s \sigma}^{\dagger} \left(\bm{e}_{i j}^{\,L} \cdot \hat{\bm{b}}_{j p \bar{\sigma}}^{\,}\right)\right]+\mathrm{h.c.}  
\end{eqnarray}
\end{widetext}
This Hamiltonian is a generalization of the models studied in Refs. \cite{Jacqmin2014,Milicevic2015,Milicevic2017,Gulevich2016,Gulevich2017,Nalitov2015,Baboux2016,Whittaker2018,Li2018} to account for the $s$-$p$ inter-orbital coupling and overlap. Here, the operator $\hat{b}_{i \ell \sigma}^{\dagger}$ ($\hat{b}_{i \ell \sigma}^{}$) creates (annihilates) a polariton at site $i$, and in its NNs, according to Eq.~(\ref{8}). The index $\ell$ labels the considered orbital in the single pillar eigenmodes ($s$, $p_{x}$ and $p_{y}$), while $\sigma=\pm$ indicates the polarization of the photon component in the circular polarization basis and $\sigma/\bar{\sigma}$ indicate opposite polarization. The coupling term for the $s$-bands, $t_s$ is spatially isotropic, while for the $p_x$ and $p_y$ orbitals (Eq.~\ref{11}) it is different depending on the orientation of the orbital with respect to the direction of the link between adjacent micropillars \cite{Wu2007,Jacqmin2014}: $t_L$ when orbitals are oriented parallel to the link, and $t_T$ for orbitals oriented perpendicular to the link (usually $|t_L|\gg| t_T|$). To describe this feature we have used a compact vector notation, similar to the one employed in \cite{Sala2013} to represent the operators that act over the $p$ orbitals, namely, 
\begin{equation}
\label{13}
\hat{\bm{b}}_{i p \sigma}=\hat{b}_{i p_{x} \sigma} \, \xv + \hat{b}_{i p_{y} \sigma}\, \yv \,.
\end{equation}
In this way the expression $(\bm{e}_{i j}^{L/T} \cdot\, \hat{\bm{b}}_{j p \sigma})$ selects the component of $\hat{\bm{b}}_{j p \sigma}$
 in the direction specified by the unit vectors
\begin{eqnarray}
\label{14a}
\bm{e}_{i j}^{\,L}&=&\cos(\varphi_{ij})\, \xv+\sin(\varphi_{ij})\, \yv \,,  \\
\label{14b}
\bm{e}_{i j}^{\,T}&=& -\sin(\varphi_{ij})\, \xv+\cos(\varphi_{ij})\, \yv \,,
\end{eqnarray}
where $L$ ($T$) indicates whether the unit vector points in the longitudinal (transverse) direction to the link $ij$, whose orientation is given by the angle $\varphi_{ij}$ (see Fig. \ref{schemes1}(a)). Therefore, these vectors select the projection of the $p_{x/y}$ orbitals parallel (perpendicular) to the lattice bond. In addition to the overlap and hopping terms that conserve the polarization we also include spin-orbit coupling (SOC) terms that flip it \cite{Nalitov2015}. For the sake of simplicity we have assumed that the SOC strength is proportional to the corresponding direct coupling, and we model it by the adimensional parameter $\Delta$. This SOC term arises from the fact that the coupling between pillars depends on whether the polariton polarization is parallel or perpendicular to the bond, owing to the fact that the two polarization modes experience different tunnel barriers in the presence of TE-TM splitting \cite{Sala2013}. 

Finally, Eq.~\ref{12} describes the coupling between $s$ and $p$ orbitals in adjacent sites, given by the coupling strength $t_{sp}$ (see Fig.~\ref{schemes1}(a)). The summation $\braket{i,j}$ in Eqs.~\ref{10}-\ref{12} runs over each site of the lattice and its NNs which is very appropriate for most lattices (see for instance Fig.~\ref{neff}(a)). An estimate of the magnitude of the second nearest neighbors hopping and overlap terms using the effective model for the photonic modes presented in the next section shows that they are roughly a factor $10^{-3}$ smaller than the ones corresponding to NNs (see also Appendix \ref{profiles}).

At this point, to describe the experimental data, all the parameters of the Hamiltonian could be taken as free fitting parameters. This has been the most common approach used in the literature so far.
While this gives reasonable results in many cases, it would be desirable to reduce the number of parameters, even though the approach might result less flexible, to gain a better physical insight and gain some predictability power on the design of different polaritonic lattices. In order to do so, we will assume in what follows that \textit{all} hopping elements $t_{\beta}$, where $\beta \in \left\lbrace s, L, T, sp \right\rbrace$, can be written as
\begin{equation}
\label{8c}
t_{\beta}=t \, S_{\beta} \, .
\end{equation}  
This is a reasonable assumption if one notices that for a system of exciton-polariton microcavity pillars spatially overlapped (see Fig.~\ref{neff}) the inter-site potential entering the definition of $t_{ij}$ (see Eq.~(\ref{3b})) might be considered as being constant within the microstructure where the two (photonic) modes mainly overlap. It is important for this to be valid to have used the symmetrized parametrization shown before so that $t_{ij}=t_{ji}$.
Equation (\ref{8c}) immediately eliminates the need to distinguish between $t_s$, $t_L$, $t_T$ or $t_{sp}$ as they are all determined by the same parameter $t$ and the corresponding overlap matrix element (which itself is fixed by the choice of a single parameter $V$ that we will introduce in the following section). 

\subsection{Single pillar mode: a simplifying approximation}
\label{approx}
Following the spirit of reducing the number of free parameters in the model, we present here a simple way to approximate the photonic eigenmodes of a single micropillar in order to calculate the overlap integrals that appear in the definition of Hamiltonian (\ref{H}). 
To this end, and for reasons that will become clear below, we consider a cylindrical microcavity defined as an infinite long ($z$-axis) circular dielectric waveguide with a step refraction index profile \cite{Gerard1996}, 
\begin{equation}
n(r) = \left\lbrace
\begin{array}{ll}
n_{1} \textup{ \,  if \,  } r\leq R, \\
n_{2} \textup{ \,  if \,  } r> R, 
\end{array}
\right.
\end{equation}
where $n_{1}$ is the refractive index of a core of radius $R$ and $n_{2}$ is the refractive index of the surrounding material. In doing this, we have ignored the $3$D nature of the problem and treat it as effectively $2$D. This is a valid assumption as far as  the wavelength of the confined modes are much larger than the micropillar resonant wavelength.  

\begin{figure}[t]
\centering
\includegraphics[width=0.95\columnwidth]{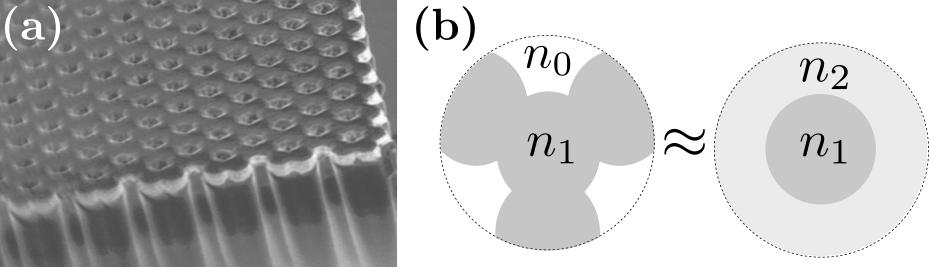}
\caption{(Color online) (a) Optical microscope photograph of a honeycomb polariton lattice where we can appreciate the spatial overlap between two adjacent micropillars. (b) Scheme of the approximation used in the Sec. \ref{approx}, in which we replace the NNs of a micropillar by an effective refraction index ($n_{2}$) in order to effectively describe the behavior of photonic modes.}
\label{neff}
\end{figure}

The calculation of the electromagnetic modes of this type of waveguide is a well-known problem in the literature (see for instance \cite{Yariv,Kapany,Gerard1996}). The exact solutions for the propagating modes are in general a mixture of transverse electric (TE) and transverse magnetic (TM) waves. Finding these hybrid modes ---usually refereed to as EH and HE modes depending on which of the electric E or magnetic H of the photon field is non-zero along the propagation direction ($z$)--- is somehow involved as it requires solving the wave equations in cylindrical coordinates with the different components of the fields being coupled. However, a good approximation for both the fields and the mode equation that determines the mode frequencies can be obtained if we assume that the core refractive index ($n_1$) is only slightly higher than that of the surrounding medium ($n_2$). This is an approximation that is often used for describing optical fibers. At first glance, this approach may seem very far from the real scenario of an isolated pillar surrounded by vacuum since $n_{1} \approx 3.5 $ and $n_{2} = n_{0} \approx 1$. However, when dealing with lattices like the one depicted in Fig.~\ref{neff}(a), where each micropillar is spatially overlapped with their NN's, it is reasonable to expect that those neighboring pillars will provide a substantially different environment to the central one, and modify the photon confinement as compared to an isolated micropillar. We propose that the presence of the neighboring pillars, which in general tends to delocalize the single pillar mode, can be effectively described by considering an effective refractive index for the surrounding medium. In this sense, our approach is variational. Because all the micropillars are made of the same material we expect $n_{2}\lesssim n_1$. Its precise value, of course, may depend on the lattice geometry: the presence of a higher/lower number of nearest neighbors will result in greater/smaller delocalization of the modes in a considered micropillar.

It is important to note at this point that even in this approach the photonic modes are well confined within the micropillar and continue to be a good starting point to build a TB model (see Appendix \ref{profiles}). Therefore, by assuming that $n_{1}-n_{2}\ll1$ we simplify the equations for matching the field components at the $r=R$ interface (see Ref.~\cite{Yariv} for details).
In this limit, the modes become linearly polarized (say, along the $\hat{x}$ and $\hat{y}$ directions), the two polarizations being  degenerated, and the electric field amplitude is given by   
\begin{equation}
\label{fields}
\nonumber
E_{lm}^{} (\bm{r}) = \left\lbrace
\begin{array}{ll}
\frac{1}{\sqrt{N}} J_{l}\left(q_{lm} r\right) \, e^{il\phi} \textup{ \, if \, } r\leq R, \\ \\
\frac{A}{\sqrt{N}} \, K_{l}\left(\kappa_{lm} r\right) \,  e^{il\phi} \textup{ \, if \, } r> R. 
\end{array}
\right.
\end{equation}
where $r$ and $\phi$ are the corresponding polar coordinates of $\bm{r}$, and $J_{l}(x)$ and $K_{l}(x)$ are the Bessel functions of the first kind and modified of the second kind, respectively. Here we have ignored the $z$-dependence of the fields (plane wave) since it is irrelevant for our purpose, $A=\frac{J_{l}\left(q_{lm} R \right)}{K_{l}\left(\kappa_{lm}R \right)}$, whereas $N$ is determined by the normalization condition. The transverse wavenumbers inside and outside of the waveguide, $\left(q_{lm}, \kappa_{lm}\right)$, are determined by the following equation
\begin{equation}
\label{trascendetalequation}
q_{lm} R\frac{J_{|l|+1}(q_{lm} R)}{J_{|l|}(q_{lm} R)}=\kappa_{lm} R \frac{K_{|l|+1}(\kappa_{lm} R)}{K_{|l|}(\kappa_{lm} R)}  \, ,
\end{equation}
where the orbital index $l = 0, \pm 1, \pm2, ...$ and the subscript $m$ indicates the $m$-th root of this transcendent equation. Equation (\ref{trascendetalequation}) can be solved numerically if one takes into account that $q_{lm}$ and $\kappa_{lm}$ are related by
\begin{eqnarray}
\label{krelations1}
n_{1}^2 k_{0}^{2} &=& q_{lm}^{2} + k_{z}^{2}\\
\label{krelations2}
n_{2}^2 k_{0}^{2} &=& k_{z}^{2} - \kappa_{lm}^{2}  
\end{eqnarray}
with $k_0=\omega_{lm}/c$ and where $\omega_{lm}$ is the frequency of the mode $lm$. It is clear then that Eqs (\ref{krelations1}) and (\ref{krelations2}) can be rewritten as 
\begin{equation}
\label{qlm}
q_{lm}R = \sqrt{V^{2}-\left(\frac{n_{2}}{n_{1}}\right)^2 \left(\kappa_{lm}R\right)^{2}},
\end{equation}
where
\begin{equation}
\label{V}
V= \frac{k_{z} R}{n_{1}} \sqrt{n_{1}^{2}-n_{2}^{2}}
=\frac{2 \pi R}{\lambda_{\text{cav}}} \sqrt{n_{1}^{2}-n_{2}^{2}}\,.
\end{equation}
Here we have used the usual experimental condition for isolated micropillars, $k_{z}=\frac{2 \pi}{\lambda_{\text{cav}}} n_{1}$, where $\lambda_{\text{cav}}$ is the micropillar resonant wavelength (see for instance Refs.~\cite{Bayer1998,Bajoni2008}).

At this point, for simplicity, we can approximate $\left(\frac{n_{2}}{n_{1}}\right)^2\approx1$ in Eq.~(\ref{qlm}), and hence the solutions of Eq.~(\ref{trascendetalequation}) can be parameterized in a very convenient way with a single parameter $V$ that effectively captures how well the electromagnetic field is confined within the micropillar. Note that in this limit, the classical electromagnetic problem is analogous to the quantum problem of a particle of mass $m$ in a finite circular potential well of radius $R$ and magnitude $U=\hbar^2V^2/2mR^2$, if we interpret the electric field as the wave function amplitude. Note also that the energies of the modes in the electromagnetic problem can be calculated as $\omega_{lm}=\frac{\hbar c}{n_{1}}\sqrt{k_{z}^2+q_{lm}^{2}}$, which in the limit $k_{z}\gg q_{lm}$ can be rewritten as
\begin{equation}
\label{w}
\omega_{lm}=\hbar c \frac{2 \pi}{\lambda_{\text{cav}}}+ \hbar c \frac{ \lambda_{\text{cav}}}{2 \pi} \frac{(q_{lm}R)^{2}}{\left(n_{1} R\right)^{2}}\,.
\end{equation}
Finally, we can define the $s$ and $p$ modes (for each polarization, linear or circular) as 
\begin{eqnarray}
\label{waveequations3}
\psi_{s }(\bm{r})&=&E_{01}(\bm{r})\,,  \\
\label{waveequations3b}
\psi_{p_{x} }(\bm{r})&=&\frac{E_{11}(\bm{r})+E_{-11}(\bm{r})}{\sqrt{2}}\,,  \\
\psi_{p_{y} }(\bm{r})&=&\frac{E_{11}(\bm{r})-\,E_{-11}(\bm{r})}{\sqrt{2}i}\,.
\end{eqnarray}
The ($2$D) overlap integrals involved in our model can be calculated as follows
\begin{eqnarray}
\nonumber
S_{s}&=&\int  \psi_{s }(\bm{r}) \psi_{s }(\bm{r}-a \xv)\, d\bm{r}\,,\\ 
\nonumber
S_{L}&=&\int  \psi_{p_{x} }(\bm{r}) \psi_{p_{x}}(\bm{r}-a \xv)\, d\bm{r}\,,\\ 
\nonumber
S_{T}&=&\int  \psi_{p_{y} }(\bm{r}) \psi_{p_{y}}(\bm{r}-a \xv)\, d\bm{r}\,,\\
\label{Integrals}
S_{sp}&=&\int  \psi_{s }(\bm{r}) \psi_{p_{x} }(\bm{r}-a \xv)\, d\bm{r}\,.  
\end{eqnarray}
We emphasize that in this approximation all the overlaps are determined by $V$, which then plays the role of a variational parameter that effectively describes the delocalization of the photonic modes due to the penetration in the adjacent overlapping micropillars as compared with those of an isolated pillar. Note that in the limit $n_{1}-n_{2}\ll1$ we are considering, the confined modes are polarization degenerate, and Eqs.~(\ref{Integrals}) are polarization independent. Polarization effects related to the spin-orbit coupling (TE-TM splitting) is phenomenologically incorporated in our model via the $\Delta$ terms in Eqs.~(\ref{10}), (\ref{11}), (\ref{12}).
\section{The honeycomb lattice\label{Results}}
The extraordinary transport and topological properties of graphene have stimulated a number of experimental and theoretical studies of the polariton honeycomb lattice \cite{Kusudo2013,Jacqmin2014,Milicevic2015,Milicevic2017,Milicevic2018,Milicevic2018a, Bleu2016,Bleu2017,Nalitov2015,Nalitov2015b,Ozawa2017a,Solnyshkov2016, Kartashov2017, Solnyshkov2018, Klembt2018}. Here we analyze the bulk band structure and the edge states spectrum  based on the complete tight-binding model presented in the previous section, highlighting the role of its different physical ingredients. We compare our numerical results with experimental data and show that they provide a very good description of the band structure. We also point out some specific signatures of the spectrum related to the photon polarization that might be relevant for future experiments. 
Finally, we reproduce recently published experimental results on the emergence of tilted Dirac cones in polariton graphene lattices under strain~\cite{Milicevic2018a}, showing that we capture correctly the dependence of parameters with distance. This gives our model certain predictive capability that could be useful to engineer different effects on artificial microcavity-polariton lattices.
\subsection{Bulk bands \label{Bulk bands}}
\begin{figure}[t]
\centering
\includegraphics[width=0.95\columnwidth]{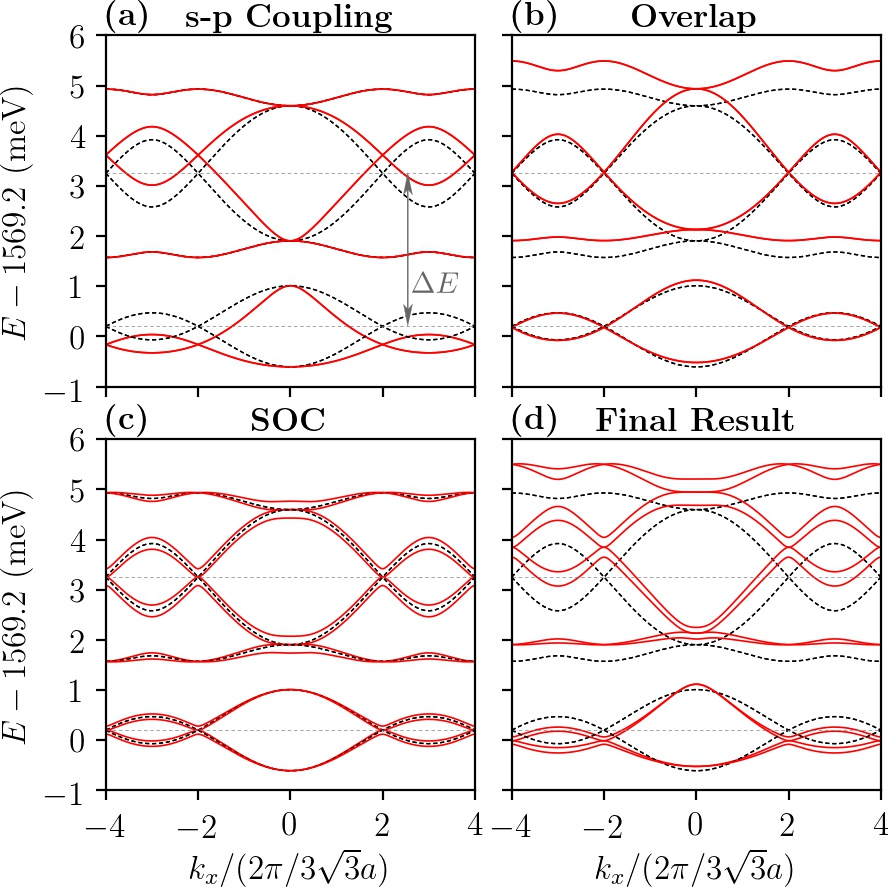}
\caption{(Color online) Calculated band structure for polariton graphene lattice as a function of the $k_{x}$ for $k_{y}=0$. Tight-binding parameters: $\Delta E=\varepsilon_{p}-\varepsilon_{s}=3.05$ meV, $t=-6.7$ meV, $\Delta=0.1$ and $V=11$. The corresponding overlap integrals are $S_{s}\approx0.04$, $S_{L}\approx-0.15$, $S_{T}\approx0.02$ and $S_{sp}\approx-0.08$ and the hoppings $t_{s}\approx0.27$ meV, $t_{L}\approx-1.01$ meV, $t_{T}\approx0.11$ meV and $t_{sp}\approx-0.52$ meV.  (a) Inter-orbital hopping effect: $t_{sp}\neq0$, $S_{\beta}=0$ and $\Delta=0$. (b) Overlap effect: $S_{\beta}\neq0$ but $S_{sp}=0$, $\Delta=0$ and $t_{sp}=0$. (c) SOC effect: $\Delta\neq0$, $S_{\beta}=0$ and $t_{sp}=0$. (d) All effects combined. The black dashed lines in all panels shows the pristine bands ($t_{sp}=0$, $S_\beta=0$, and $\Delta=0$). }
\label{effects}
\end{figure}

To achieve a better understanding of the influence of the different terms in Hamiltonian (\ref{H}) let us consider first the bulk bands along a specific direction of high symmetry of the underlying lattice. For that, we  define the lattice vectors  $\bm{a}_{1}  =  \sqrt{3} \,  a \, \xv $, $ \bm{a}_{2} =  \frac{\sqrt{3}}{2} \,  a \,  (\xv+ \sqrt{3} \,  \yv)$ and the relative position of the basis sites $A$ and $B$,  $\bm{\delta}_{AB}  =  \frac{a}{2}\,( \sqrt{3}\, \xv+ \yv)$ (see Fig. \ref{schemes1}(b)). Here $a$ is the distance between two NNs pillars. The calculated spectrum for a lattice of micropillars of a diameter $D = 2R=3~\mu$m and a center-to-center distance $a=2.4~\mu$m, along $k_{y}=0$,  is show in the Fig.~\ref{effects}. In the different panels of the figure, we analyze the contribution of the different terms of the model  separately, that is, considering only one of them at a time. In each panel, the black dashed line represents the bands in the absence of $s$-$p$ coupling ($t_{sp}=0$), non-orthogonality ($S_\beta=0$), and SOC ($\Delta=0$). In this case each band is clearly particle-hole symmetric and the upper and lowermost $p$-bands present a very small dispersion --this bands would be completely flat for $t_T=0$. The red solid lines in each panel of Fig.~\ref{effects}(a)-(c) include each contribution separately: (a) only $s$-$p$ coupling ($\Delta=0$, $S_{\beta}=0$), (b) only non-orthogonality ($\Delta=0$, $S_{sp}=0$ and $t_{sp}=0$), and (c) only SOC ($S_{\beta}=0$, $t_{sp}=0$) --see the figure caption for the value of all the parameters. 

As clearly seen in the figure, each term  leads to a different effect on the bands.  The inter-orbital coupling  (Fig.~\ref{effects}(a)) plays a very important role on the deformation of the bands as it tends to join them, stretching the top of the $s$-band and the bottom of the $p$-band, and making them to acquire a V-like shape in the neighborhood of the $\Gamma$ point. Notice also that the uppermost and lowermost $p$-bands are not affected by this coupling. This is to be expected as those bands involved the $p$-orbitals that are perpendicular to the bonds and hence they do not couple to the $s$-bands.
On the contrary, one of the main effects of the non-orthogonality between orbitals in different sites (Fig.~\ref{effects}(b)) is to produce a clear asymmetry between those quasi-flat bands, making the uppermost wider and the lowermost narrower. This point is quite relevant as it will allow us to reproduce the experimental data quite well without the need to include an energy-dependent hopping (as previously done in ~\cite{Jacqmin2014}). We can highlight that the non-orthogonality induces the opposite asymmetry of the bands as compared to the $s$-$p$ coupling. This effect is significant for the $p$-bands, while it remains negligible for the $s$-bands. Indeed, from the overlap integrals Eq.~(\ref{Integrals}) we estimate the overlap between $p$ orbitals in adjacent micropillars to be of the order of $15$\% for typical lattices, while it is only $4$\% for the $s$-bands.
The effect of the SOC (Fig.~\ref{effects}(c)) is simply to split the bands, as expected, leading to the appearance of a polarization (spin) texture. Finally, Fig.~\ref{effects}(d) shows the bands including all terms. 
\subsection{A comment on the effect of non-orthogonality\label{nonorthogonality}}
We would like to emphasize some important aspect of our model for cavity polariton lattices. On the one hand, we neglect the second nearest neighbors ($2$NN) hopping terms. This is so because in most of the lattices there is no overlap between $2$NNs micropillars and hence the coupling of the photonic modes goes through the extremely weak evanescent field in vacuum, out of the micropillars. 
On the other hand, we do include the non-orthogonality between orbitals located in adjacent sites. Its effect on the asymmetry of the bulk bands (Fig.~\ref{effects}(b)) could be qualitatively reproduced in a phenomenological way by including an effective $2$NN hopping term (see, for instance, Ref.~\cite{Jacqmin2014}). However, $2$NN hopping would have a very different effect on the flat band states localized in zigzag and bearded edges: the $2$NNs hopping would destroy the flatness of the edge states band, while the non-orthogonality preserves it. 
This can be easily understood as follows. We re-write Eq.~(\ref{2}) as 
\begin{equation}
\label{eq_overlap_effect}
 \bm{H}\bm{c}=\varepsilon\, \bm{ c}+\varepsilon\, \tilde{\bm{S}}\bm{ c},
\end{equation}
where $\tilde{\bm{S}}$ is the non-diagonal part of $\bm{S}$. Now, we will assume, for the sake of simplicity and to make the argument clear, that we are only considering a set of equivalent orbitals so that all energy sites can be taken to be equal to zero without any loss of generality. In addition, we continue to assume that the  hopping terms are proportional to the overlap between NNs sites. Under these assumptions $\tilde{\bm{S}}=\lambda\bm{H}$, with $\lambda<0$ and $|\lambda|\ll1$. Therefore
\begin{equation}
\label{eq_overlap_effect2}
 \bm{H}\bm{c}=\frac{\varepsilon}{1-\lambda\, \varepsilon}\, \bm{ c}\,,
\end{equation}
and so the band structure is given by
\begin{equation}
\label{eq_overlap_effect3}
\varepsilon_{\bm{k}}=\frac{\bar{\varepsilon}_{\bm{k}}}{1+\lambda\, \bar{\varepsilon}_{\bm{k}}}\, ,
\end{equation}
with $\bar{\varepsilon}_{\bm{k}}$ the energy dispersion of the orthogonal case ($\lambda=0$). It is then clear that: \textit{i}) flat bands remain flat when the non-orthogonality is included. In particular, the ones at $\bar{\varepsilon}_{\bm{k}}=0$ do not move when non-orthogonality is included. Therefore, flat band edge states characteristics of zigzag and armchair edges are not affected by non orthogonal effects; \textit{ii}) upper (lower) bands that correspond to $\bar{\varepsilon}_{\bm{k}}>0$ ($\bar{\varepsilon}_{\bm{k}}<0$) gets broader (narrower) as the factor $(1+\lambda\, \bar{\varepsilon}_{\bm{k}})^{-1}$ is bigger (smaller) than $1$. Of course, in addition to this there is also some deformation of the original band. We emphasize once again that this effect is opposite to the one induced by the $s$-$p$ coupling (see Fig.~\ref{effects}(b)). Therefore, the presence in the experimental data of a clear asymmetry between the lowest and uppermost $p$-bands is an evidence of the importance of the non-orthogonality while the asymmetry of the inner middle bands is an indication of the relevance of the $s$-$p$ coupling. 

When more orbitals are involved and coupled, as in our numerics, the above picture gives only a qualitative description of the non-orthogonality effect. Furthermore, we expect this picture to hold even if we extend our model and include (slightly) different proportionality constants between the different hopping terms and the corresponding overlap matrix elements.
\subsection{Polariton graphene ribbon and comparison with the experiment\label{Polariton Graphene ribbon}}
We now calculate the band structure for a polariton graphene ribbon (PGR) with zigzag edges (see Fig.~\ref{schemes1}(b)), taken to be  infinite along the $x$ direction (hence we can use Bloch theorem) and containing $30$ unit cells along the transverse direction (defined by $\bm{a}_2$). We use the same TB parameters as above. 

To analyze the polarization properties of the bulk (center of the ribbon) and edge states, and compare with the data obtained from photoluminescence experiments, we define a quantity that describes the probability of detecting a state (via the emission of a photon) located at the site $\mathbf{R}_{n \alpha}=n \mathbf{a}_{2}+\bm{\delta}_{\alpha}$ (where $\bm{\delta}_{\alpha}$ gives the position of the non-equivalent micropillar $\alpha \in \left\lbrace A,B\right\rbrace$ in the unit cell) with polarization $\sigma$ and  quasi-momentum $\bm{k}=k_x\,\hat{\bm{x}}+k_y\,\hat{\bm{y}}$. Namely, 
\begin{equation}
\label{prob}
p_{n \alpha\bm{k}}^{\sigma}=\frac{1}{C}\sum_{\ell', \alpha'}\left|  \sum_{n'} e^{-i n' \bm{k}\cdot\bm{a_{2}}}  G_{n' \alpha' n \alpha} (\bm{S}_{k_{x}} \bm{c}_{k_{x}})_{n'\alpha'\ell' \sigma}\right|^{2},
\end{equation}
where $G_{n' \alpha' n \alpha}=\exp\left[\frac{-\left[\left(\mathbf{R}_{n' \alpha'}-\mathbf{R}_{n \alpha}\right).\yv\right]^2}{2 \Sigma^2}\right]$ is a Gaussian weight function focused on the site $\mathbf{R}_{n \alpha}$ with standard deviation $\Sigma$, which allows to replicate the spatial dependence of the emitted light in experiments. Indeed, in photoluminescence experiments of photonic graphene micropillars, the escape of photons out of the microcavity results in a Gaussian-like distribution of the emission centered at the position of the pump spot~\cite{Jacqmin2014}. $\bm{S}_{k_{x}}$ and $\bm{c}_{k_{x}}$ are the ribbon overlap matrix and eigenvectors,  respectively (see Appendix \ref{Hribbon}), and $C$ is determined for each $k_{x}$ by normalization condition.  

\begin{figure}[t]
\centering
\includegraphics[width=0.95\columnwidth]{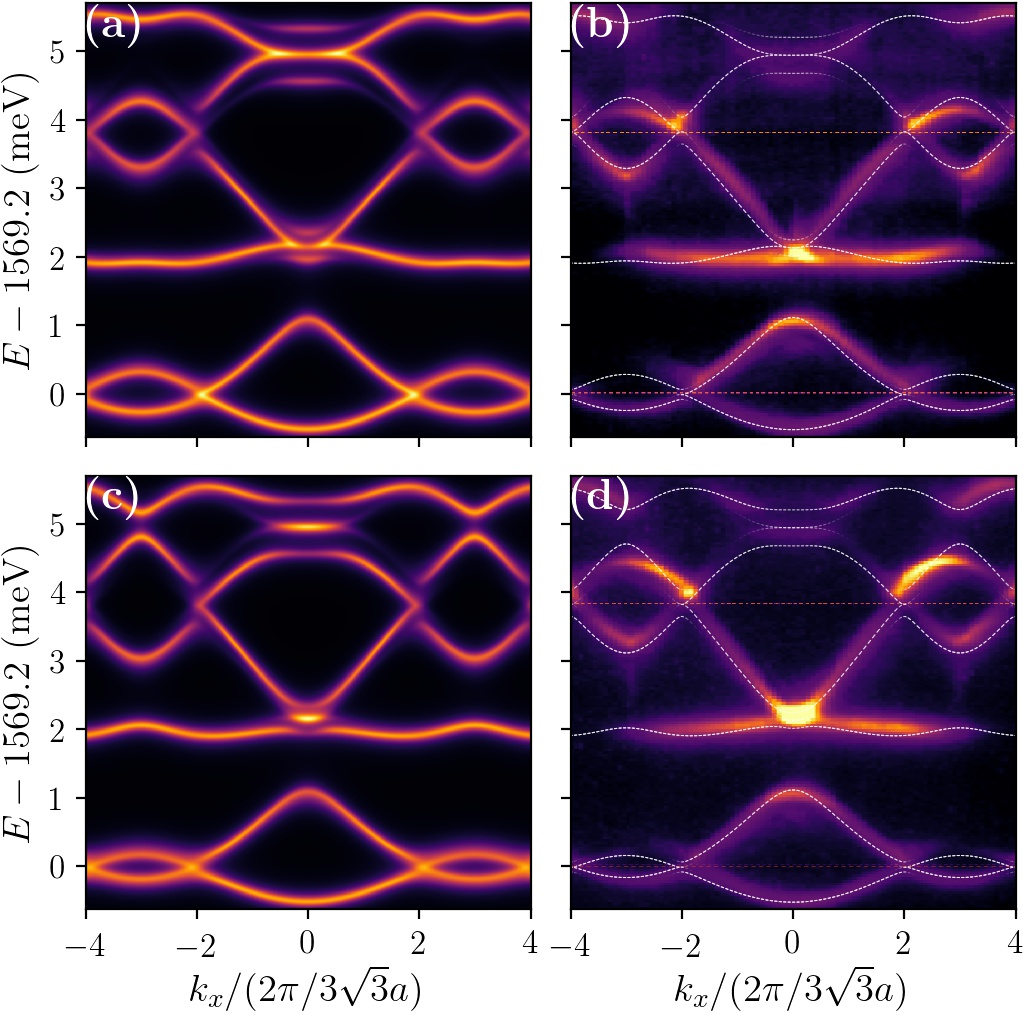}
\caption{(Color online) Bulk emission spectrum for a PGR with zigzag edges as a function of the wave vector parallel to the edges (with $k_{y}=4\pi/3a$, equivalent to $k_{y}=0$). (a)-(c) Calculated spectrum for a polarization along the edge ($\sigma=\parallel$) and perpendicular to it ($\sigma=\perp$), respectively.  The color scale of each panel has been normalized to its maximum value. TB parameters: $\Delta E=3.05$ meV, $t=-6.7$ meV, $\Delta=0.1$ and $V=11$.
(b)-(d) Corresponding experimental data. In dot white lines we reproduce the polarization calculated bands for a bulk system (Fig.~\ref{effects}(d)) for the purpose of comparison.}
\label{expbulk}
\end{figure} 
\begin{figure*}[t]
\begin{center}
\includegraphics[width=0.95\columnwidth]{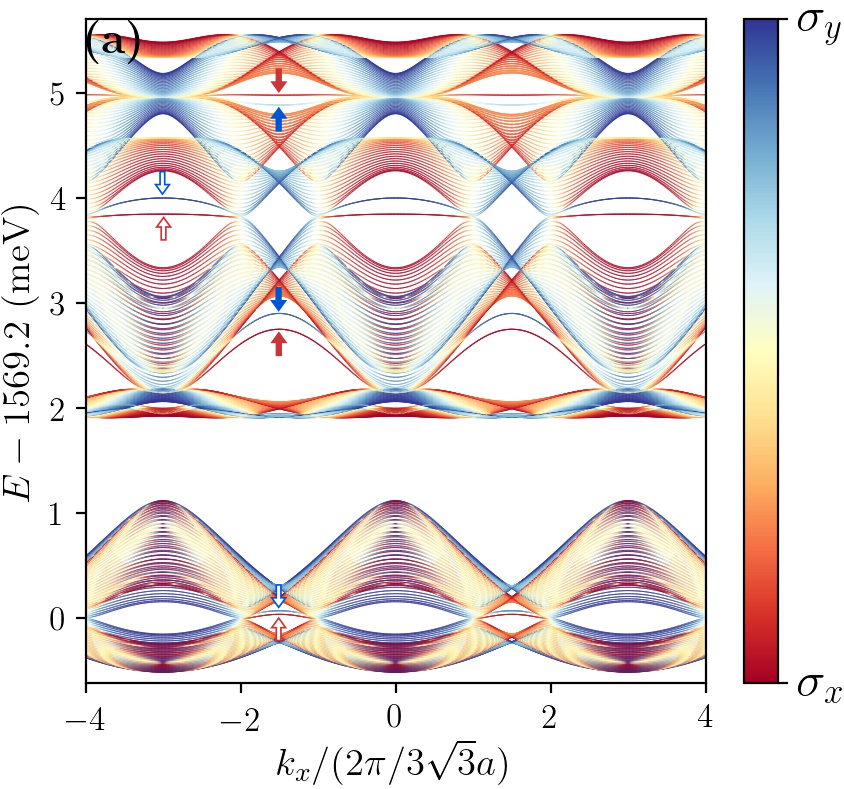}
\includegraphics[width=0.95\columnwidth]{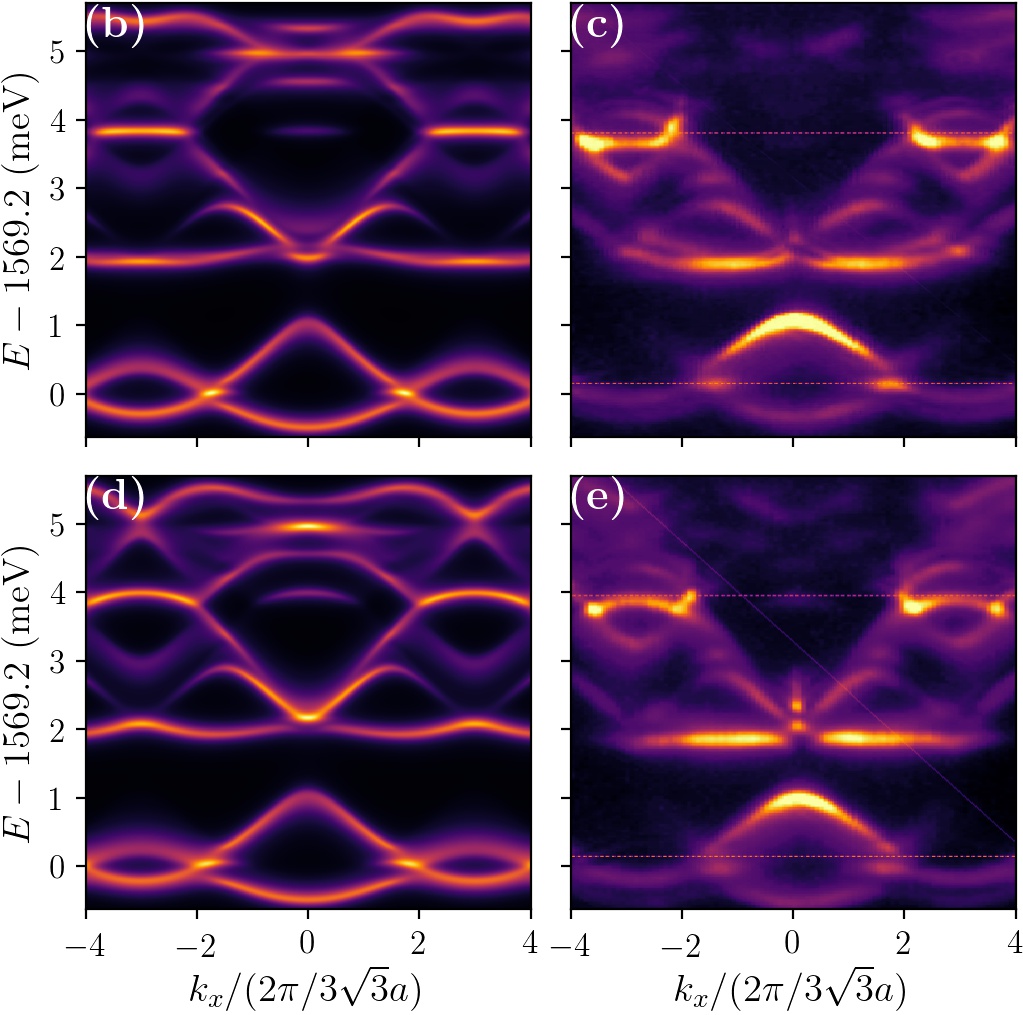}
\end{center}
\caption{(Color online) (a) Calculated band structure for a zigzag PGR. The TB parameters are the same as in the previous figure, except that the $p_y$ orbitals at the edge pillar have a different energy ($\delta\varepsilon'_{p_{y}\parallel}=0.4$ meV and $\delta\varepsilon'_{p_{y}\perp}=0.60$ meV, see text). The color scale encodes the polarization of the states. Red and blue correspond to the states polarized along the $x$ ($\parallel$) and $y$ ($\perp$) axis, respectively. Edge states are highlighted with arrows (solid and open arrows indicated different types of edge states, see text). (b) and (d) Same as Fig. \ref{expbulk} but projecting in a site located at the edge of the ribbon. (c) and (e) Corresponding experimental data.}
\label{edge}
\end{figure*} 
Figures \ref{expbulk}(a) and \ref{expbulk}(c) show the calculated emission spectrum ($p_{n \alpha\bm{k}}^{\sigma}$) for the case of excitation of a bulk site located at the center of the ribbon, $\left(n=15, \alpha=A\right)$--- along the path $\bm{k}=k_x\,\hat{\bm{x}}$, and setting $\Sigma=3a$. We included in the simulations an artificial broadening parameter ($\gamma=0.07$ meV) with the sole purpose of reproducing the experimental linewidth. Figures~\ref{expbulk}(b) and (d), show the corresponding experimental results for the photon emission polarized parallel or perpendicular to the ribbon's edge, respectively. The experimental conditions are those of Ref.~\onlinecite{Milicevic2017}: an AlGaAs-based microcavity, with 28(40) Bragg pairs in the upper(lower) mirror, with 12 GaAs quantum wells etched into a lattice of micropillars of a diameter of $3~\mu$m and a center-to-center distance $a$ of $2.4~\mu$m; a non-resonant laser at 740~nm excites the bulk of the lattice in a spot of $4~\mu$m in diameter; the light emitted from the polariton bands is collected as a function of the linear polarization direction, emitted angle (in-plane momentum) and wavelength. In Fig.~\ref{expbulk} the bands are measured along the $k_x$ direction for an angle of emission in the $y$ direction corresponding to $k_{y}=4\pi/3a$, passing through the centre of the second Brillouin zone. The reason for selecting the emission through the second Brillouin zone is to avoid destructive interference effects characteristics of bipartite lattices that prevent clear observation of the bands at the center of the first Brillouin zone~\cite{Shirley1995,Jacqmin2014}.
In order to make a detailed comparison between experiment and theory, in Fig.~\ref{expbulk}(b) and (d), on top of the experimental data we plot the numerical results (dotted lines) corresponding to the bulk system polarized in the $x$ or $y$ direction as appropriate (same data as in Fig.~\ref{effects}(d)). Many details of the experimental spectrum  are clearly captured by our simple model. We emphasize that our approach for this (bulk) case involves only a few fitting parameters: $\Delta E$, $t$, $V$ and $\Delta$.  

The corresponding figures for the case of photoluminescence from a site at the edge of the ribbon ($n=1, \alpha=A$) are presented in Fig.~\ref{edge}. 
Here the spectrum is slightly more complex as several new features appear so a more detailed analysis is needed.
Figure~\ref{edge}(a) shows the complete ribbon band spectrum calculated with our TB model. Red and blue lines correspond to the parallel ($\parallel$) and perpendicular ($\perp$) polarization with respect to the ribbon edge, respectively. Localized edge states (indicated by the arrows) appear both at the $s$ and the $p$ bands. In the latter case, there are two types of edge states, as discussed in Ref.~ \cite{Milicevic2017}: (i) the usual edge states (open arrows), similar to the ones on the $s$ bands, that appear near the Dirac cones and are usually flat---here the dispersion observed for one of the polarizations is mainly due to the difference on the onsite energy of the edge site, see discussion below---, and (ii) the ones with a non-trivial dispersion (solid arrows). In the latter case the splitting is caused mainly by the inclusion of the SOC coupling, being the bands rather polarized.

Figures \ref{edge}(b) and \ref{edge}(d) show the corresponding calculated spectrum ($p_{1 Ak_x}^{\sigma}$)  while Figs. \ref{edge}(c) and \ref{edge}(e) shows the measured bands along the $k_x$ direction for an angle of emission in the $y$ direction corresponding to $k_{y}=4\pi/3a$. 
A careful analysis of the latter shows that : (i)  there is a clear difference  between both polarizations in the case of the `flat' edge states (the Dirac cone edge states, highlighted with open arrows in Fig.~\ref{edge}(a)): polarization parallel to the edge (Fig.~\ref{edge}(c)) shows a flat edge state (as naively expected) while for the perpendicular polarization (Fig.~\ref{edge}(e)) it is dispersive
---we emphasize here that this effect cannot be accounted for by the inclusion of a second NNs hopping as the later is negligible;
(ii) the SOC induced splitting of the `dispersive' edge states (those highlighted with solid arrows in Fig.~\ref{edge}(a)) is a bit stronger for the lower edge bands as compared with the upper edge bands.

We have found that these features can be accounted for in our model by modifying the site energy of the surface (edge) pillars as compared with those of the bulk orbitals. In particular, only the energy of the $p_y$ orbital needs to be modified, being different for each polarization. Hence, all the results shown in Fig.~\ref{edge} include such a change, which is  given by $\delta\varepsilon'_{p_{y}\parallel}=0.4$ meV and $\delta\varepsilon'_{p_{y}\perp}=0.6$ meV.
A possible origin for this energy shift might be the combination of the presence of excitonic stress at the edge pillars of the lattice and different confinement of photonic modes when the number of nearest neighbors is reduced with respect to the bulk pillars (micropillars in the zigzag edge have two NNs while in the bulk all pillars have three NNs).

\subsection{Strain induced merging of Dirac cones \label{merging}}

\begin{figure}[t]
\centering
\includegraphics[width=0.95\columnwidth]{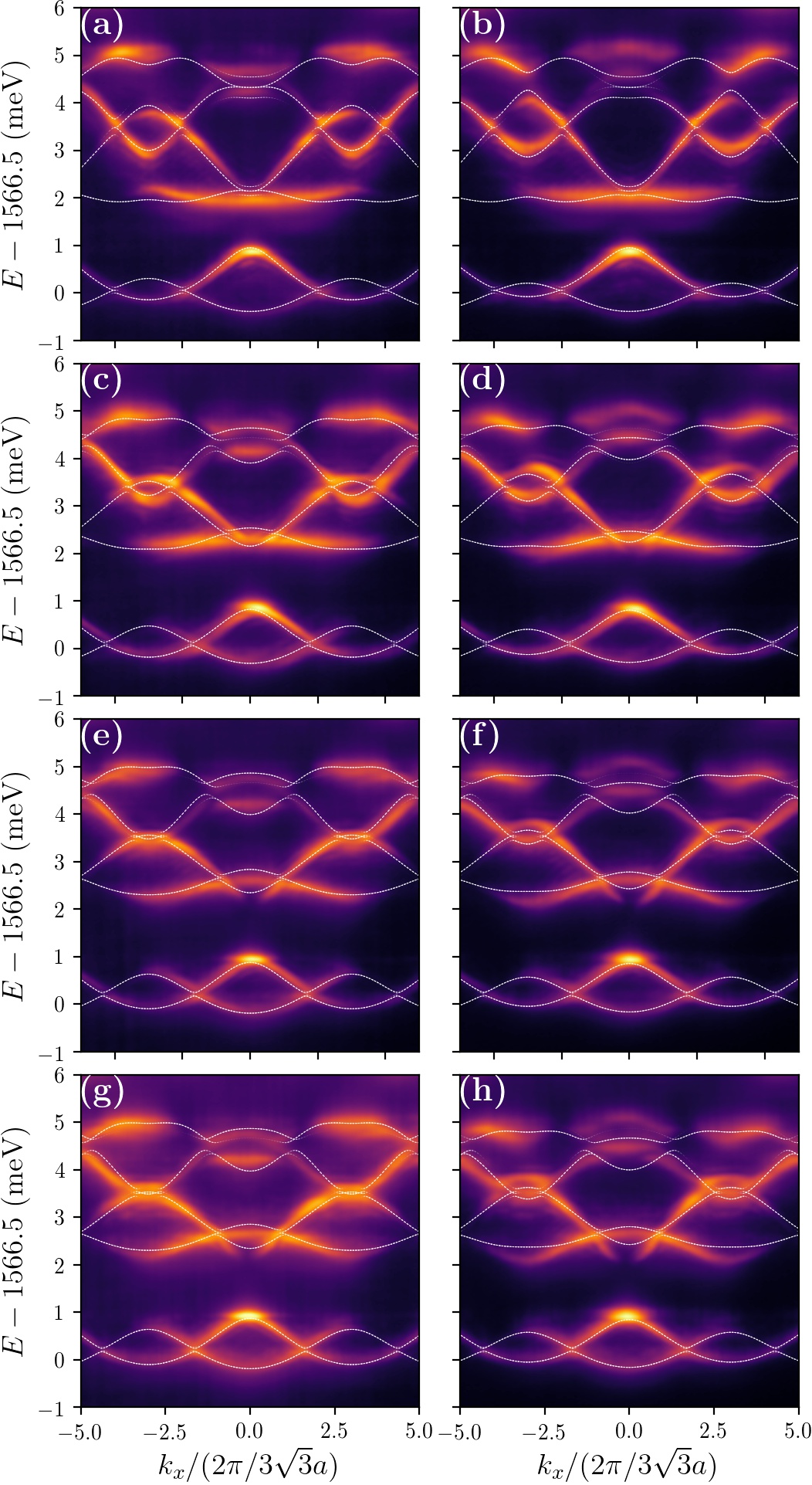}
\caption{(Color online) Measured polariton photoluminescence intensity as a function of $k_{x}$ for different values of $a'$ (the colour scale of each panel has been independently normalized to its maximum value). The cut is done for $k_{y} = 2 \pi/(a'+a/2)$. Left/right column, polarization parallel/perpendicular to the edge. (a)-(b) Unstrained ribbon $a'=2.40\mu$m. Strained ribbons: (c)-(d) $a'=2.60\mu$m, (e)-(f) $a'=2.70\mu$m and (g)-(h) $a'=2.72\mu$m.  In dashed white lines we reproduce the polarized calculated bands for a corresponding bulk system. Tight-binding parameters: $t=-5.0$ meV, $\Delta=0.1$, $V=6.5$ and for $a'=2.40\mu$m, $\Delta E=2.81$ meV; $a'=2.60\mu$m, $\Delta E=2.90$ meV and for $a'=2.70\mu$m and $a'=2.72\mu$m, $\Delta E=3.00$ meV.}
\label{transicion}
\end{figure}

One of our goals in developing this tight binding model is not only to account correctly for all the different couplings and non-orthogonality effects, and establish their relative importance, but also to be able to predict the band structure up to some fine details as this would be very useful in the design of future experiments.
To show this potentiality, we have analyze the case of a distorted honeycomb lattice as the one used in Ref.~\cite{Milicevic2018a}, where the length of the bond perpendicular to the zigzag edge was changed to be $a'\geq a$. 
Figure \ref{transicion} shows experimental luminescence measured at the center of the lattices under similar conditions as Fig.~\ref{edge}, at an exciton photon detuning at the bottom of the $p$ bands of $-10$~meV. Here the micropillars are $2.75~\mu$m in diameter and a center-to-center distance of the undistorted bonds of $a=2.4~\mu$m, with the strained bond being $a'=2.4~\mu$m (undistorted lattice, (a)-(b)), $2.6~\mu$m (c)-(d), $2.7~\mu$m (e)-(f) and $2.72~\mu$m (g)-(h). The left/right column corresponds to the emission linearly polarized parallel/perpendicular to the edge. Solid white lines show the calculated dispersions.
 All the parameters of our model (except for $\Delta E$, which was slightly corrected for each distance) where changed only through their dependence with the bond distance. Quite notably, in agreement with the experimental data, we find that the merging of the Dirac cones occurs for $a'=2.7$ $\mu$m for the parallel polarization---for the perpendicular polarization we observe a similar behavior but for a larger distortion due to the presence of the SOC. The latter is a signal that the magnitudes of the overlap and SOC are correct.  
\section{Final remarks\label{conclu}}
We have presented a relatively simple tight-binding model to describe generic cavity polariton lattices including  the most relevant physical ingredients. Namely, the coupling between single pillar modes of different symmetry ($s$-$p$ coupling) and the non-orthogonality between different sites. A careful analysis and comparison with the experimental data allowed us to identify the most prominent features each contribution introduces and, although they change the band structure with similar magnitudes, it turns out that the $s$-$p$ coupling leads to the most distinguishable effects. This coupling substantially reshapes the bands, particularly the $s$-band, a feature so far neglected in experimental and theoretical polariton studies.

The non-orthogonality plays an important role in the $p$-bands, resulting in an asymmetry in the dispersion of the uppermost and lowermost $p$-bands. Our estimates show that the $p$ orbitals have an overlap between adjacent micropillars of the order of $15$\% --for typical lattices--, while for the $s$ ones it is only $4$\% and, hence, non-orthogonal effects can be safely ignored for the $s$-bands.

In concordance with this, it is important to emphasize that second NNs hopping is negligible as there is essentially no overlap between second nearest micropillars, and the evanescent field out of the etched micropillars decreases extremely fast. Note that this might not be the case in polariton lattices fabricated with other techniques. In particular, lattices fabricated by partial etching of the structure (upper mirror)~\cite{Klembt2017, Whittaker2018}, metallic deposition on the surface, or intracavity mesa techniques ~\cite{Kaitouni2006,Winkler2015}, might present deeper evanescent fields and may result in significant second NNs couplings.
We stress, however, that second NN hopping and non-orthogonality act very differently  on the flat band states localized in zigzag and bearded edges: while the former destroys the flatness of the edge states band, the latter  preserves it.

While here we restricted ourselves to consider only the $s$ and $p$ modes, it is rather natural to ask whether higher energy modes should also be included. Calculations show that by adding the $d$ modes similar results are obtained. However, comparisons with experiments reveal that, although qualitatively the structure of the $d$ bands is well captured by the model, it overestimates its bandwidth and its coupling with the lower modes. That is, the measurements show a greater confinement for the $d$ bands than the expected for the model. We argue that this may be due to the proximity of $d$ bands to the exciton-energy, which in the experiments shown here amounts to $-10$ meV for the $p$ bands and about to 0 for the $d$ bands, much closer to the exciton resonance. Therefore,  the excitonic contribution to the $d$ polariton states is greater than for the lower modes and 
the photonic component smaller, then reducing the hopping between pillars and correspondingly the bandwidth.  
Preliminary calculations using a different parameter $V'$ for $d$ bands show slightly better description of the experiment. Yet, that comes at the price of increasing the number of parameters of our simplified parameterization with very few fitting parameters and it does not result in a relevant improvement for the $s$ and $p$ bands.
 
The theoretical results here presented provide accurate guidelines to describe the band structure of lattices of polariton micropillars, and explain the break up of the particle-hole symmetry observed experimentally and assigned, up to now, to second nearest neighbors effects.

\acknowledgements
We acknowledge financial support from ANPCyT (grants PICTs 2013-1045 and 2016-0791), from CONICET (grant PIP 11220150100506), from SeCyT-UNCuyo (grant 06/C526), the ERC grant Honeypol, the Quantera grant Inerpol, the FETFLAG grant PhoQus, the French National Research Agency (ANR) project Quantum Fluids of Light (ANR-16-CE30-0021), the Labex CEMPI (ANR-11-LABX-0007) and NanoSaclay (ICQOQS, Grant No. ANR-10-LABX-0035), the French RENATECH network, the CPER Photonics for Society P4S, the I-Site ULNE via the project NONTOP and the Métropole Européenne de Lille via the project TFlight. GU acknowledges support from the ICTP associateship program and 
thanks the Simons Foundation.
\appendix
\section{Wave functions profiles and penetration length\label{profiles}}
\begin{figure}[t]
\centering
\includegraphics[width=.9\columnwidth]{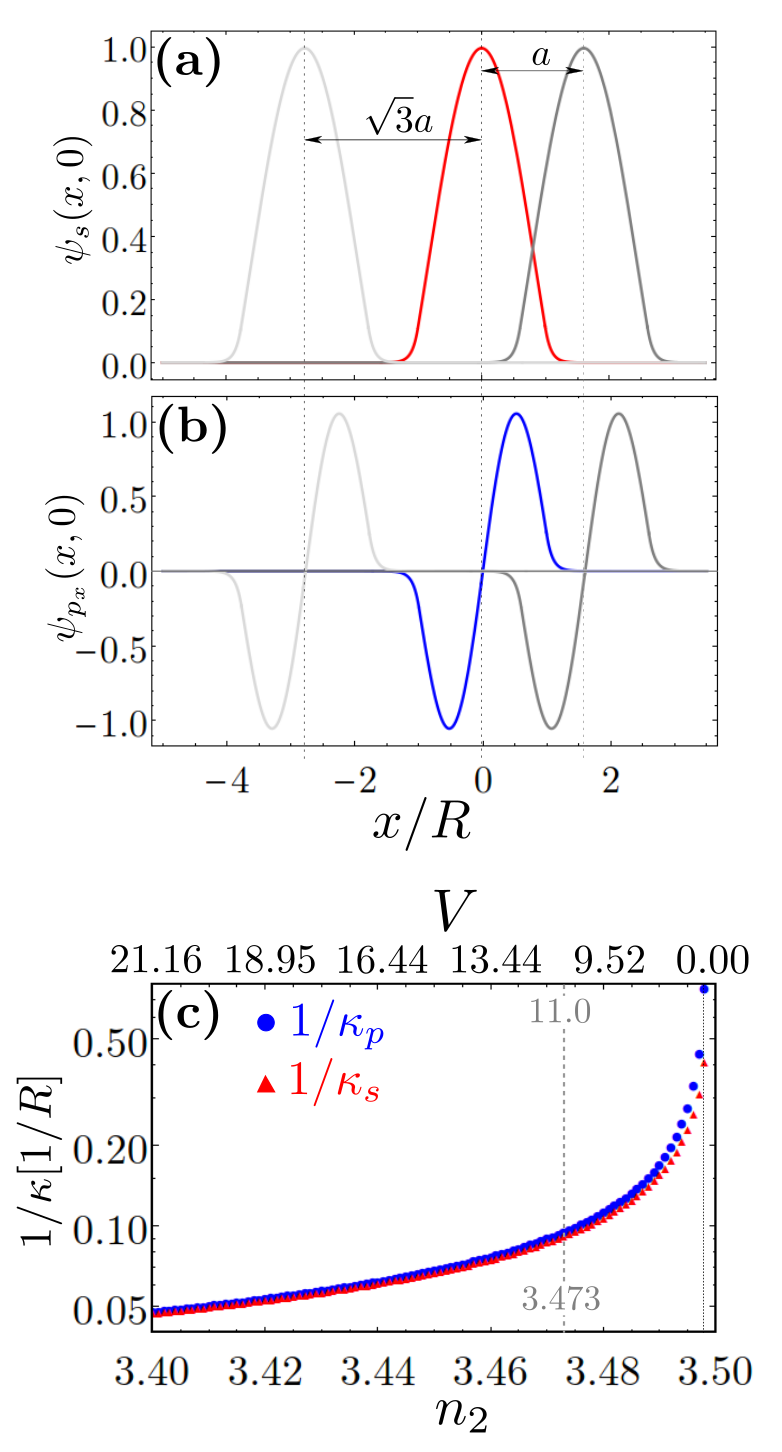}
\caption{(Color online) (a)-(b) Spatial profiles of the $s$ and $p$ modes, respectively. We show three functions in each case, centred on $x=0$, $x = a$ (NNs distance for honeycomb lattice) and  $x = \sqrt{3} a$ (NNNs distance for honeycomb lattice). (c) Penetration length calculated as $\kappa^{-1}$ for $n_{2} \in [3.400,3.498]$.}
\label{prof}
\end{figure}

The approach presented in Sec. \ref{approx} might be, at first glance, somehow anti-intuitive about how well confined is the wave function within the micropillar. To try to clarify this point we show in Fig.~\ref{prof}(a) and \ref{prof}(b) the profiles of the wave functions $\psi_{s}(x,0)$ and $\psi_{p_{}x}(x,0)$, respectively. In each case we plot the  wave function centered at $x=0$,  $x = a$ (NNs distance for the honeycomb lattice) and $x = \sqrt{3} a$ (NNNs distance for the honeycomb lattice). We use the similar parameters as in the main text, $a = 2.40$ $ \mu$m, $D=2R = 3.00 ~\mu$m and $V=11.0$. Using these values, and approximate experimental values for the micropillar refraction index ($n_{1} \approx 3.5$) and the resonance wavelength of the cavity ($\lambda_{\text{cav}}\approx 740$ nm) we calculate from Eq.~(\ref{V}) the value of the effective refractive index of the external medium $n_{2}\approx 3.4733$. As we see in the figures, although $n_{2}$ has a value very close to $n_{1}$, the wave functions are still well confined, so the values of the NNs overlap integrals are small $S_{s}\approx0.04$, $S_{L}\approx-0.15$, $S_{T}\approx0.02$ and $S_{sp}\approx-0.08$ while for NNNs these integrals are negligible $S_{s}\approx3\times 10^{-5}$, $S_{L}\approx-2\times 10^{-4}$, $S_{T}\approx1\times 10^{-5}$ and $S_{sp}\approx-9\times 10^{-5}$. In addition,  we show the penetration length for the $s$ and $p$ modes (Fig. \ref{prof}(c)) defined as $1/\kappa_{s}$ and $1/\kappa_{p}$, respectively, for $n_{2} \in [3.400,3.498]$---here $\kappa_{s}=\kappa_{01}$ and $\kappa_{p}=\kappa_{11}$, see Eqs. (\ref{waveequations3}) and (\ref{waveequations3b}). Notably,  the penetration lengths are very small even for values of $n_{2}$ very close to $n_{1}$. In the case of the $s$ mode $\kappa_{s}^{-1} \propto e^{\frac{2}{V^{2}}} $  when $V \rightarrow 0$ ($n_{2} \rightarrow n_1$). This non-analytic behavior is rather particular for a confinement potential in $2$D. For the $p$ modes the behavior is slightly more complex and $\kappa_{p}^{-1}\rightarrow \infty$ when $V \rightarrow z_{01}$ ($n_{2} \rightarrow n_{c}=3.49831$)---here $z_{01}$ is the first zero of the $J_0$ Bessel function. The fact that the penetration length diverges for a finite value of the potential  means that for $n_2>n_{c}$ the $p$ modes are not confined in the micropillar.   

\section{Two couple pillars: effective media approximation}
\begin{figure}[t]
\centering
\includegraphics[width=0.95\columnwidth]{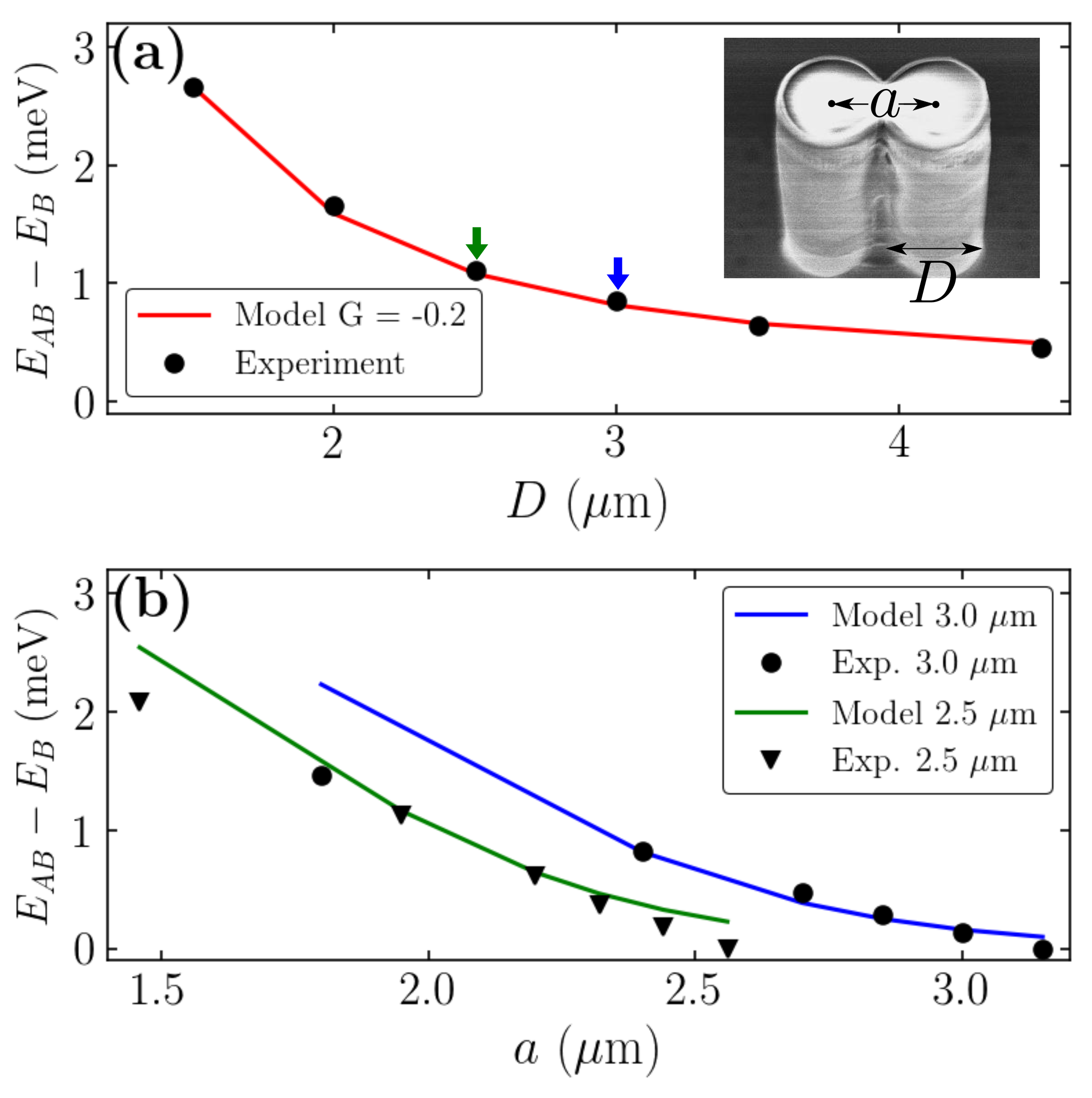}
\caption{(Color online)(a) Measured (extract from \cite{Vasconcellos2011}) and calculated energy splitting of the first two optical modes of the PM for various diameters keeping $G=-0.20$. (b) The same as (a) but as a function of the pillar's distance keeping the radius constant (note that in this case the value of G changes).}
\label{molecule}
\end{figure}

To test the range of validity of our approach we consider here the case of two coupled pillars, the so-called polaritonic molecule (PM) \cite{Vasconcellos2011}, as those shown in Fig.~\ref{molecule}(a). We notice that this is not the best scenario for our variational approximation to the single micropillar photonic mode as the effective index $n_2$ used to represent the surrounding pillars is somehow harder to justify in this configuration. Yet, we will show that even in this case it provides a very satisfying phenomenological description of the energy separation between the first two modes of the PM, the bonding (B) and anti-bonding (AB) modes. 

We use the Hamiltonian (\ref{H}), without considering the effect of SOC ($\Delta=0$), to fit the experimental results \cite{Vasconcellos2011} for the splitting ($E_{AB}-E_{B}$) as a function of diameter ($D = 2R$) of the micropillars. This is done in Fig. \ref{molecule}(a) keeping the parameter $G=a/D-1$ constant. Here $a$  is the separation between the centers of two micropillars. This corresponds to maintain the normalized overlap between the pillars unchanged. Note that for $G < 0$, $G=0$, $G > 0$, the micropillars overlap, are tangent, and are separated, respectively. Note that to fit correctly the experiment we have to consider the $R$ dependence of the parameters $V$ and $\Delta E$. Namely, from Eq.~(\ref{V})  we have
\begin{equation}
V = v R,
\end{equation}
and from Eq.~(\ref{w})  
\begin{equation}
\Delta E = \frac{\delta E}{R^{2}}\,,
\end{equation}
where $v$ and $\delta E$ where taken as  adjusting parameters.
The results for the model using $t=-5.2$ meV, $v=2.1 \frac{1}{\mu \text{m}}$ and $\Delta E = 40.5 \frac{\text{meV}}{\mu \text{m}^{2}}$ and the experiment value $G=-0.20$ are shown in  Fig.~\ref{molecule}(a). The agreement with the experimental data is very good. Using the same value of the parameters we show in Fig. \ref{molecule}(b) the comparison with the experimental splitting ($E_{AB}-E_{B}$) as a function of the distance between the centers of the micropillars , keeping $D$ constant (that is, modifying the value of $G$), for $D=2.5 \mu$m and $D=3.0 \mu$m. In this case we can see a good agreement between model and experiment for values of $G$ between $0$ and $-0.25$. The discrepancies observed for $G>0$ are expected since for this condition the pillars are separated and the model loses validity. In the other case, for $G<-0.25$, the discrepancies can be understood by noting that the approximation given by Eq.~(\ref{8c}) overestimates the value of the real hopping integral. 
\section{Ribbon Hamiltonian \label{Hribbon}}
When considering an infinite long ribbon along the $x$ direction, with a finite width ($y$ direction), it is better decompose the eigenstates in plane waves along the ribbon's direction and define a crystal momentum $k_{x}$. In this case the eigenstates of the system can be written as
\begin{equation}
 \ket{\Psi_{k_{x}}}=\sum_{n=1}^{N_{n}}\sum_{\alpha \ell \sigma} c_{k_{x} n \alpha \ell \sigma} \ket{\psi_{k_{x} n \alpha \ell \sigma}}\,,
\end{equation}
where the Bloch wave functions are given by 
\begin{equation}
\ket{\psi_{k_{x} n \alpha \ell \sigma}}=\sum_{m=1}^{N_{m}} e^{i m k_{x} a_{1x}} \ket{\psi_{m n \alpha \ell \sigma}}\, .
\end{equation}
Here $m$ is the index that lists the transverse layers that make up the ribbon and $n\alpha$ is a composite index that labels the intra layer elements,  so that the position of each micropillar is given by
$\left(m\, \bm{a}_{1}+n\, \bm{a}_{2}\right)+\bm{\delta}_{\alpha}$
where $\bm{a}_{1}=a_{1x}\xv$ and $\bm{a}_{2}$ are the primitive vectors, while $\bm{\delta}_{\alpha}$ gives the position of the non-equivalent micropillar $\alpha \in \left\lbrace A,B\right\rbrace$ in the unit cell. The other two indices, $\ell$ and $\sigma$, refer to the orbital and polarization degrees of freedom, respectively.

As mentioned in Sec. \ref{Non-orthogonal tight-binding approach}, the problem is then reduced to solving the matrix equation 
\begin{equation}
 \bm{H}_{k_{x}} \bm{c}_{k_{x}}=\varepsilon_{k_{x}} \bm{S}_{k_{x}}\bm{c}_{k_{x}}\,,
\end{equation}
where in this case the ribbon Hamiltonian and overlap matrix can be written as 
\bea
\nonumber
\bm{H}_{k_{x}}&=&\left(\bm{H}^{(0)}+e^{i k_{x} a_{1x}}\bm{H}^{(+1)}e^{-i k_{x} a_{1x}}\bm{H}^{(-1)}\right)\,,\\
\label{17}
\bm{S}_{k_{x}}&=&\left(\bm{S}^{(0)}+e^{i k_{x} a_{1x}}\bm{S}^{(1)}+e^{-i k_{x} a_{1x}}\bm{S}^{(-1)}\right)\,.
\eea
The matrix elements of the layer matrices defined above are given by
\bea
\nonumber
H_{n \alpha \ell \sigma n' \alpha' \ell' \sigma'}^{(m')}&=&\bra{\psi_{0 n \alpha \ell}}\hat{H} \ket{\psi_{m' n' \alpha' \ell' \sigma'}}\,,\\
S_{n \alpha \ell \sigma n' \alpha' \ell' \sigma'}^{(m')}&=&\langle\psi_{0 n \alpha \ell}\ket{\psi_{m' n' \alpha' \ell' \sigma'}}\,.
\eea

%

\end{document}